\documentclass[11pt]{article}
\usepackage{graphicx}
\usepackage{amssymb}
\usepackage{amsmath}
\usepackage{times}
\usepackage{helvet}
\usepackage{courier}
\usepackage{cite}
\newcommand{\be}{\begin{eqnarray}}\newcommand{\beq}{\begin{equation}}
\newcommand{\ee}{\end{eqnarray}}\newcommand{\eeq}{\end{equation}}

\input epsf 
\input amssym.def 
\input amssym.tex 

\numberwithin{equation}{section}

\title{Geometric view of the thermodynamics of adsorption at a line of three-phase contact}
\author{Y. Djikaev and B. Widom \\ \\ Department of Chemistry, Baker Laboratory, Cornell University\\ Ithaca, New York  14853-1301}

\begin{document}
\date{}
\maketitle

We consider three fluid phases meeting at a line of common contact and study the linear excesses per unit length of the
contact line (the linear adsorptions $\Lambda_i$) of the fluid's components.  In any plane perpendicular to the contact
line, the locus of choices for the otherwise arbitrary location of that line that makes one of the linear adsorptions,
say $\Lambda_2$, vanish, is a rectangular hyperbola.  Two of the adsorptions, $\Lambda_2$ and $\Lambda_3$, then both
vanish when the contact line is chosen to pass through any of the intersections of the two corresponding hyperbolas
$\Lambda_2 = 0$ and $\Lambda_3 = 0$.  There may be two or four such real intersections.  It is required, and is confirmed
by numerical examples, that a certain expression containing $\Lambda_{1(2,3)}$, the adsorption of component 1 in a frame
of reference in which the adsorptions $\Lambda_2$ and $\Lambda_3$ are both 0, is independent of which of the two or four
intersections of $\Lambda_2 = 0$ and $\Lambda_3 = 0$ is chosen for the location of the contact line.  That is not true of
$\Lambda_{1(2,3)}$ by itself; while the adsorptions and the line tension together satisfy a linear analog of the Gibbs
adsorption equation, there are additional, not previously anticipated terms in the relation that are required by the line
tension's invariance to the arbitrary choice of location of the contact line.  The presence of the additional terms is
confirmed and their origin clarified in a mean-field density-functional model.  The additional terms vanish at a wetting
transition, where one of the contact angles goes to 0.
\newpage 
\section{Introduction} 
It was long ago remarked by Buff \cite{Buff1964} (as quoted by Melrose \cite{Melrose1968}
and Good \cite{Good1976}), and Lovett \cite{Lovett1965}, that a measurement of the temperature or chemical-potential
dependence of the interfacial tension between two phases can yield a length characteristic of the microscopic structure
of the interface; viz., the distance between two well-defined Gibbs dividing surfaces.  It is then a microscopic length
obtainable from a macroscopic measurement, which is remarkable.  What makes it possible is that the measurement is made
in three dimensions while what is measured is a property of the interface, and thus of a system of one less dimension. 
The result is then the characteristic length of the ``missing" dimension, which is microscopic.  It is a similar
reduction in dimension from three to two that made possible Rayleigh's famous estimate of the length of a molecule by
measurement of the area of a monolayer that had spread from a known volume \cite{Rayleigh1899}.

The purpose of the present paper is to give a geometric interpretation of the temperature or chemical-potential
dependence of the tension of the line of contact of three phases, analogous to that found by Buff and Lovett for the
interface between two phases.  Here we first rederive the result of Buff and Lovett, and we then state the analogous
question for a three-phase contact line.

Let the tension of the interface between two phases $\alpha$ and $\beta$ be $\sigma_{\alpha\beta}$, the chemical
potentials of the various chemical components $i$ of the system be $\mu_i$, and the densities of those components in the
bulk phases be $\rho_i^\alpha$ and $\rho_i^\beta$.  It is convenient \cite{Rowlinson1982}, for simplicity of notation, to
include the temperature $T$ among the $\mu_i$ and the entropy densities $s^\alpha$ and $s^\beta$ among the
$\rho_i^\alpha$ and $\rho_i^\beta$, so that in a c-component system the index $i$ = 1, 2,\ldots, $c$+1.

In Fig.\:1 is shown, schematically, the two phases $\alpha$ and $\beta$ with a diffuse interface between them,
and an arbitrary Gibbs dividing surface chosen to be at the height $z_0$ on the vertical scale of distance, $z$.  Let
$\Gamma_i$ be the adsorption (surface excess per unit area) of the component $i$ with respect to that dividing surface. 
Then the Gibbs adsorption equation is 
\begin{equation}  \label{eq:1} d\sigma_{\alpha\beta} = - \sum\limits_{i}
~\Gamma_{i} ~d\mu_{i} . \end{equation} 
Of the $c+1$ chemical potentials $\mu_i$ (among which is the temperature $T$) only
$c$ are independent because they are related by the Clapeyron equation [6] 
\begin{equation} \label{eq:2} \sum\limits_{i}
~(\rho_i^\alpha - \rho_i^\beta)~d\mu_i ~= 0. \end{equation} 
\indent The adsorption $\Gamma_{i}(z)$ depends on the
location $z_0$ of the dividing surface via 
\begin{equation} \label{eq:3} \Delta ~ \Gamma_i = (\rho_i^\alpha -
\rho_i^\beta) ~\Delta z \end{equation} 
where $\Delta\Gamma_i$ is the change in $\Gamma_i$ when the dividing surface is
displaced by the distance $\Delta z$ in the direction from phase $\beta$ towards phase $\alpha$ \cite{Buff1964,
Melrose1968, Good1976, Lovett1965, Rowlinson1982}.  Let $z_i$ and $z_j$ be the locations for the dividing surface that
make $\Gamma_i = 0$ and $\Gamma_j = 0$, respectively, and, in standard notation, let $\Gamma_{i(j)}$ be the adsorption
$\Gamma_i$ when the dividing surface is chosen to be at $z_j$, where $\Gamma_j = 0$.  Then from (1.3),  \begin{equation}
\label{eq:4} \Gamma_{i(j)} = ~(\rho_i^\alpha - \rho_i^\beta)~(z_j - z_i). \end{equation}

In a one-component system, or in a c-component system at fixed $\mu_3,\ldots, \mu_{\rm c+1}$, there is only one
independently variable $\mu_i$, say $\mu_1$.  Then from (1.1) and (1.4), \begin{equation} \label{eq:5} \frac
{d\sigma_{\alpha\beta}}{d\mu_1} =~(\rho_1^\alpha - \rho_1^\beta)~(z_1 - z_2) ~~~(\rm fixed~ \mu_3,\ldots, \mu_{c+1});
\end{equation} i.e., from a measurement of $d\sigma_{\alpha\beta}/d\mu_1$ and the known densities in the bulk phases one
obtains both the magnitude and sign of the distance $z_1 - z_2$ between the two dividing surfaces of $\Gamma_1 = 0$ and
$\Gamma_2 = 0$.  This is the Buff-Lovett principle \cite{Buff1964, Melrose1968, Good1976, Lovett1965, Rowlinson1982}.

In Fig.\:2 we show three phases $\alpha, \beta, \gamma$ meeting in a line of common contact, which is where the three
interfaces $\alpha\beta, ~\beta\gamma$, and $\alpha\gamma$ meet.  We do not consider curvature effects, so the interfaces
are planar and the contact line is straight.  The dihedral angles between interfaces are the contact angles, which we
also call $\alpha, \beta$, and $\gamma$, with $\alpha + \beta + \gamma = 2 \pi$.  The location, although not the
direction, of the contact line is arbitrary, just as is the location of the dividing surface in Fig.\:1.  In Fig.\:3 is a
view along the contact line, which is perpendicular to the plane of the figure, so it appears as a point, the interfaces
appearing as lines meeting in that point.  Two alternative and equally arbitrary choices for the location of the contact
line, and so also for the locations of the interfaces, are shown in the figure.  There is thus a two-dimensional infinity
of choices for the location of the contact line; i.e., it may be chosen to pass through any point in the plane of the
figure.  In the analogous Fig.\:1 for the interface between two phases there is only a one-dimensional infinity of
choices of dividing surface.  The line tension $\tau$, like the surface tensions $\sigma_{\alpha\beta}$, etc., is
independent of what choice is made.

There are linear adsorptions $\Lambda_i$ (linear excesses of the components $i$ per unit length of the contact line)
analogous to the surface adsorptions $\Gamma_i$, and there are curves in the plane of Fig. 3 that are the loci of the
choices for the location of the contact line for which any of the linear adsorptions $\Lambda_i$ is 0.  We ask what those
loci are, and what geometrical information is contained in $\Lambda_{1(2,3)}$, the linear adsorption of component 1 in a
frame of reference in which $\Lambda_2 = \Lambda_3 = 0$; i.e., when the contact line is chosen to be at the intersection
of the curves $\Lambda_2 = 0$ and $\Lambda_3 = 0$.  Likewise, we ask for the geometric interpretation of $d\tau/d\mu_1$,
and of the connection between $d\tau/d\mu_1$ and $\Lambda_{1(2,3)}$.  The answers will be the analogs for the three-phase
contact line of the relations (1.1), (1.4), and (1.5) for two-phase interfaces, but will prove to be a more complicated
story.

In Section 2 we find that the locus of choices for the contact line that make $\Lambda_i = 0$ (briefly, the curve
$\Lambda_i = 0$) is a rectangular hyperbola, and that any two such hyberbolas, say $\Lambda_2 = 0$ and $\Lambda_3 = 0$,
may intersect in two or four points.  We find, however, that $\Lambda_{1(2,3)}$ depends on which of the intersections is
chosen as the location of the contact line.  This means that $-\Lambda_{1(2,3)}$ cannot be $d\tau/d\mu_1$, which is
invariant to the arbitrary location of the contact line; i.e., that the linear analog of the Gibbs adsorption equation
(1.1) must contain hitherto unrecognized terms that restore the invariance.  This is the subject of Section 3.  The
additional terms are found to vanish at a wetting transition, which is where one of the contact angles closes down to 0. 
In Section 4 we confirm in a simple mean-field density-functional description what in Section 3 we suggest is the
physical origin of the extra terms, and we see how in principle these terms would be calculable from a microscopic
theory.  The results are briefly summarized in the concluding Section 5.

\section{Linear adsorption $\Lambda_i$, locus $\Lambda_i = 0$, and $\Lambda_{i(j,k)}$}
In Fig. 4 we show the Neumann triangle associated with the three-phase equilibrium.  This is in the plane of Fig. 3, with
a particular choice of location of the contact line.  The sides of the triangle are perpendicular to the interfaces
$\alpha\beta, ~\beta\gamma$, and $\alpha\gamma$, intersecting them at the respective distances $R_{\alpha\beta}$,
$R_{\beta\gamma}$, and $R_{\alpha\gamma}$ from the contact line.  The vertex angles of the triangle are the supplements
$\pi-\alpha$ etc. of the contact angles.  The three interfaces divide the triangle into the three areas $A_\alpha,
A_\beta, A_\gamma$, as shown.

We introduce the unit vectors $\bold{e}_{\alpha\beta}, \bold{e}_{\beta\gamma}, \bold{e}_{\alpha\gamma}$ in the directions
of the interfaces, and another set of unit vectors $\bold{g}_{\alpha\beta}, \bold{g}_{\beta\gamma},
\bold{g}_{\alpha\gamma}$ that are respectively perpendicular to the $\bold{e}_{\alpha\beta}$, etc., and are rotated from
them counterclockwise.  These are illustrated in Fig. 5.  There will later also occur dyads of the form
$\bold{e}_{\alpha\beta}~ \bold{g}_{\alpha\beta}$, etc.

With the Neumann triangle fixed in space, the distances $R_{\alpha\beta}(\bold{r})$ etc., areas $A_{\alpha}(\bold{r})$
etc., and surface adsorptions $\Gamma_i^{\alpha\beta}(\bold{r})$ etc., are all functions of the arbitrary location
$\bold{r}$ of the contact line.  For two different choices $\bold{r}_0$ and $\bold{r}$ they are such that
\begin{equation} \label{eq:6} R_{\alpha\beta}(\bold{r}) - R_{\alpha\beta}(\bold{r}_0) = -\bold{e}_{\alpha\beta} \cdot
(\bold{r}-\bold{r}_0) , \end{equation} 
and similarly for $R_{\beta\gamma}$ and $R_{\alpha\gamma}$; 
\begin{align}
\label{eq:7} A_{\alpha}(\bold{r}) - A_{\alpha}(\bold{r}_0) = &-R_{\alpha\beta}(\bold{r}_0)~\bold{g}_{\alpha\beta} \cdot
(\bold{r}-\bold{r}_0) + R_{\alpha\gamma}(\bold{r}_0)~\bold{g}_{\alpha\gamma} \cdot (\bold{r}-\bold{r}_0) \nonumber \\
&+\frac{1}{2} (\bold{r}-\bold{r}_0) \cdot (\bold{e}_{\alpha\beta}~\bold{g}_{\alpha\beta} -
\bold{e}_{\alpha\gamma}~\bold{g}_{\alpha\gamma}) \cdot (\bold{r}-\bold{r}_0),  \end{align} 
and similarly for $A_\beta$ and $A_\gamma$ with $\alpha, \beta, \gamma$ permuted cyclically; and, from (1.3), 
\begin{equation} \label{eq:8}
\Gamma_i^{\alpha\beta}(\bold{r}) - \Gamma_i^{\alpha\beta}(\bold{r}_0) = (\rho_i^\alpha -
\rho_i^\beta)~\bold{g}_{\alpha\beta} \cdot (\bold{r}-\bold{r}_0), \end{equation} 
and similarly for
$\Gamma_i^{\beta\gamma}$ and $\Gamma_i^{\alpha\gamma}$ with $\alpha, \beta, \gamma$ permuted cyclically.  Note that these
are invariant to the interchange of $\bold{r}$ and $\bold{r}_0$.  That is obvious for (2.1) and (2.3), and for (2.2) it
is a consequence of (2.1).  Note also from (2.2) and its analogs for $A_\beta$ and $A_\gamma$ obtained from $A_\alpha$ by
cyclic permutation of $\alpha, \beta, \gamma$ that $A_\alpha + A_\beta + A_\gamma$ is independent of the location of the
contact line, as it must be because it is just the total area of the Neumann triangle.

It is to be appreciated that while the contact line may be located arbitrarily at any $\bold{r}$ in the plane of Fig.\:4,
the properties of the system are spatially varying and, with the physical system itself fixed in space, different choices
of $\bold{r}$ correspond to physically different structures about the contact line.  While
$\Gamma_i^{\alpha\beta}(\bold{r})$ does vary with $\bold{r}$, it is a well defined and in principle measurable function
of it.

That is true also of the linear adsorptions $\Lambda_i$, which are functions $\Lambda_i(\bold{r})$ of the choice of
location $\bold{r}$ of the contact line with respect to which they are defined.  For any two such choices $\bold{r}_0$
and $\bold{r}$,
\begin{equation} \label{eq:9}
	\begin{split} 
	   \Lambda_i({\bf r}) - \Lambda_i({\bf r}_0) = - \left\{ \rho_i^{\alpha} \left[ A_{\alpha} ({\bf r})\right.\right. 
		& \left. - ~A_{\alpha} ({\bf r}_0) \right] + \cdots \\ 
 		&  + \big[
		\Gamma_i^{\alpha\beta} ({\bf r})\;R_{\alpha\beta}({\bf r}) - \Gamma_i^{\alpha\beta} ({\bf r}_0)
		~R_{\alpha\beta}({\bf r}_0)\big] + \cdots \left.\right\}\end{split}  
\end{equation}
where the omitted terms marked by $\cdots$ in the first line have $\beta$ and $\gamma$ in place of $\alpha$ and those in
the second line have $\beta\gamma$ and $\alpha\gamma$ in place of $\alpha\beta$.  From (2.1)-(2.4), then, 
\begin{equation} \label{eq:10} \begin{split} 
\Lambda_i({\bf r}) - \Lambda_i({\bf r}_0) = \big[
\Gamma_i^{\alpha\beta}({\bf r}_0) {\bf e}_{\alpha\beta} + \Gamma_i^{\beta\gamma}({\bf r}_0) {\bf e}_{\beta\gamma} +
\Gamma_i^{\alpha\gamma}({\bf r}_0) {\bf e}_{\alpha\gamma} \big] \cdot ({\bf r} - {\bf r}_0) \\ 
+ ~\frac{1}{2} ({\bf r}-{\bf r}_0) 
\cdot \big[(\rho_i^\alpha - \rho_i^\beta){\bf e}_{\alpha\beta} ~{\bf g}_{\alpha\beta} + (\rho_i^{\beta} -
\rho_i^{\gamma}) {\bf e}_{\beta\gamma} ~{\bf g}_{\beta\gamma} \big.\\ 
 \big. + ~(\rho_i^{\gamma} - \rho_i^{\alpha}) {\bf e}_{\alpha\gamma} ~{\bf g}_{\alpha\gamma}\big] \cdot ({\bf r}-{\bf r}_0). 
\end{split} \end{equation} 
Note that there is no remaining reference to the distances $R_{\alpha\beta}$ etc.: the difference $\Lambda_i({\bf
r})-\Lambda_i({\bf r}_0)$ depends only on the locations ${\bf r}_{0}$ and $\bf r$ in the physical system and makes no
reference to the Neumann triangle.  Note also that (2.5) is invariant to the interchange or $\bf r$ and ${\bf r}_{0}$, as
a consequence of (2.3).

	When the ``component" $i$ is the grand-canonical free energy, its linear adsorption $\Lambda_i$ is the line tension $\tau$, its surface adsorptions $\Gamma_i^{\alpha\beta}$ etc. are the interfacial tensions $\sigma_{\alpha\beta}$ etc., and its bulk-phase densities $\rho_i^\alpha, \rho_i^\beta, \rho_i^\gamma$ are all equal to $-p$, the negative of the uniform bulk-phase pressure (since the interfaces are assumed planar and the contact line straight).  Then (2.5) together with the condition
\begin{equation} \label{eq:11}
\sigma_{\alpha\beta}~{\bf e_{\alpha\beta}} + \sigma_{\beta\gamma}~{\bf e_{\beta\gamma}} + \sigma_{\alpha\gamma}~{\bf e_{\alpha\gamma}} = 0
\end{equation}
of mechanical equilibrium yields $\tau({\bf r}) - \tau({\bf r}_0) = 0$, which is the required invariance of the line tension to the location of the contact line.

	If we have a system of rectangular coordinates $x,y$ with origin at ${\bf r}_{0}$ and with the $x$ and $y$ axes in the directions of $\bf e_{\alpha\gamma}$ and $\bf g_{\alpha\gamma}$ respectively, the equation $\Lambda_i({\bf r}) = 0$, from (2.5), is the conic
\begin{equation} \label{eq:12}
A_iy^2 + B_ixy + C_ix^2 + D_iy + E_ix + F_i = 0
\end{equation}
with
\begin{align} \label{eq:13}
-C_i &= A_i = -\frac{1}{4} (\rho_i^\alpha - \rho_i^\beta) \sin 2\alpha + \frac{1}{4} (\rho_i^\beta - \rho_i^\gamma) \sin 2\gamma \nonumber \\
B_i &= \frac{1}{2} (\rho_i^\gamma - \rho_i^\alpha) + \frac{1}{2} (\rho_i^\alpha - \rho_i^\beta) \cos 2\alpha + \frac{1}{2} (\rho_i^\beta - \rho_i^\gamma) \cos 2\gamma \nonumber \\
D_i &= -\Gamma_i^{\alpha\beta} ({\bf r}_0) \sin \alpha + \Gamma_i^{\beta\gamma} ({\bf r}_0) \sin \gamma \nonumber \\
E_i &= \Gamma_i^{\alpha\beta} ({\bf r}_0) \cos \alpha + \Gamma_i^{\beta\gamma} ({\bf r}_0) \cos \gamma + \Gamma_i^{\alpha\gamma} ({\bf r}_0) \nonumber \\
F_i &= \Lambda_i ({\bf r}_0),
\end{align}
where $\alpha$ and $\gamma$ are two of the contact angles:  $\bf e_{\alpha\beta} \cdot \bf e_{\alpha\gamma} = \rm cos~
\alpha$ and $\bf e_{\beta\gamma} \cdot \bf e_{\alpha\gamma} = \rm cos~ \gamma$.  Because $C_i = -A_i$ the conic is a
rectangular hyperbola (perpendicular asymptotes).  Although the form of its equation depends on the choice of origin
${\bf r}_0$ and the chosen orientation of the $x$ and $y$ axes, the hyperbola itself does not; it is a physically
well-defined curve with a well-defined location and orientation in any plane perpendicular to the direction of the
contact line, and could in principle (although perhaps only with difficulty in practice) be traced by measurements on the
fixed equilibrium system.

Let $\Lambda_j({\bf r}) = 0$ be similarly defined as the locus of those choices $\bf r$ for the location of the contact
line that make the linear adsorption $\Lambda_j = 0$.  The two hyperbolas $\Lambda_i({\bf r}) = 0$ and $\Lambda_j({\bf
r}) = 0$ may in general make two or four real intersections with each other.  They cannot make more than four, as can be
seen by translating and rotating axes so that one of the hyperbolas becomes of the form $xy = constant$ in the new
coordinate system.  Then substituting $y = constant/x$ in the equation of the other and multiplying through by $x^2$
yields a polynomial equation in $x$ of the fourth degree for the locations of their intersections, which can have at most
four real roots.  No intersections, or confluent intersections, are possible but non-generic.

	 On dividing (2.7) through by $F_i$ and recalling that $C_i = -A_i$,  we see that the hyperbola $\Lambda_i({\bf r}) = 0$ is determined by four parameters, $A_i/F_i , ~B_i/F_i, ~D_i/F_i$, and $E_i/F_i$.  Thus, typically only one such hyperbola, not two distinct ones, could pass through four given points, but two distinct ones may do so if the four points in question, $(x_1, y_1), \ldots , (x_4, y_4)$, are related by
\begin{equation} \label{eq:14}
 \left| \begin{array} {cccc} x_1^2 - y_1^2 ~&~ x_1y_1 ~&~ x_1 ~&~ y_1  \\
\\
 x_2^2 - y_2^2 ~&~ x_2y_2 ~&~ x_2 ~&~ y_2  \\
\\
 x_3^2 - y_3^2 ~&~ x_3y_3 ~&~ x_3 ~&~ y_3  \\
 \\
 x_4^2 - y_4^2 ~&~ x_4y_4 ~&~ x_4 ~&~ y_4  \end{array} \right| = 0~~,
\end{equation}
for then the four points would no longer determine the four parameters $A_i/F_i$ etc. uniquely. 
Thus, either the two rectangular hyperbolas $\Lambda_i({\bf r}) = 0$ and $\Lambda_j({\bf r}) = 0$ intersect each other in two points, or in four, and in the latter case the four intersections satisfy (2.9).

	Figure 6 illustrates the case of two intersections of the hyperbolas $\Lambda_2 = 0$ and $\Lambda_3 = 0$, with the parameters
\begin{align} \label{eq:15}
A_2/F_2 = -1, ~~B_2/F_2 = 9/10, ~~D_2/F_2 = E_2/F_2 = 3  \nonumber \\
A_3/F_3 = 0, ~~B_3/F_3 = -1/5, ~~~D_3/F_3 = E_3/F_3 = 0  ~.
\end{align}
Figure 7 illustrates the case of four intersections, with
\begin{align} \label{eq:16}
A_2/F_2 = -1, ~~ B_2/F_2 = D_2/F_2 = E_2/F_2 = 0 \nonumber \\
A_3/F_3 = -1/10, ~~ B_3/F_3 = -1/5, ~~D_3/F_3 = 2/5, ~~E_3/F_3 = -1  ~.
\end{align}
Values of the parameters on the right-hand sides of (2.8) that are implied by these values of $A_i/F_i$ etc. are in the Appendix, as are numerically accurate values for the $x,y$ coordinates of the various intersections.  The coordinates of the four intersections in Fig.\:7 satisfy (2.9).

	Now let ${\bf r}_{ jk}$ be any one of the two or four intersections of $\Lambda_j({\bf r}) = 0$ with
$\Lambda_k({\bf r}) = 0$ and let ${\bf r}_{i}$ be any point on the locus $\Lambda_i({\bf r}) = 0$.  Then from (2.5),
taking $\bf r$ to be ${\bf r}_{i}$ and ${\bf r}_{0}$ to be ${\bf r}_{jk}$, we obtain for $\Lambda_{i(j,k)}$, the linear
adsorption $\Lambda_i$ in a frame of reference in which the contact line is chosen to be at a point where the linear
adsorptions $\Lambda_j$ and $\Lambda_k$ are both 0,
\begin{equation} \label{eq:17} 
	\begin{split} 
\Lambda_{i(j,k)}  = - \big[ \Gamma_i^{\alpha\beta}({\bf r}_{jk}) {\bf
e}_{\alpha\beta} + \Gamma_i^{\beta\gamma}({\bf r}_{jk}) {\bf e}_{\beta\gamma} + \Gamma_i^{\alpha\gamma}({\bf r}_{jk})
{\bf e}_{\alpha\gamma}\big] \cdot ({\bf r}_i - {\bf r}_{jk}) \\ 
- ~\frac{1}{2} ({\bf r}_i-{\bf r}_{jk}) \cdot
\big[(\rho_i^\alpha - \rho_i^\beta){\bf e}_{\alpha\beta} ~{\bf g}_{\alpha\beta} + (\rho_i^\beta - \rho_i^\gamma) 
{\bf e}_{\beta\gamma} ~{\bf g}_{\beta\gamma} \big. \\ 
\big. + ~(\rho_i^\gamma - \rho_i^\alpha) {\bf e}_{\alpha\gamma} ~{\bf
g}_{\alpha\gamma} \big] \cdot ({\bf r}_i-{\bf r}_{jk}) . 
	\end{split} 
\end{equation}
This can be written in several alternative ways by taking advantage of the invariance of (2.5) to the interchange of $\bf
r$ and ${\bf r}_0$, but the form in (2.12) is as convenient as any.

Equation (2.5), or the nearly equivalent (2.12), is fundamental in our analysis.  Equation (2.5) is the analog
for the contact line of (1.3) for an interface and (2.12) is the analog of (1.4).  In the following section we consider
the linear analogs for the line tension of the Gibbs adsorption equation (1.1) and the Buff-Lovett principle (1.5), and
the roles of $\Lambda_i$ and $\Lambda_{i(j,k)}$ in these relations.

\maketitle
\section{Gibbs adsorption equation for the contact line}

One reads in Gibbs \cite{Gibbs1928}, `{`\it{These}} {\rm[three-phase contact]} {\it{lines might be treated in a manner
entirely analogous to that in which we have treated surfaces of discontinuity.  We might recognize linear densities of
energy, of entropy, and of the several substances which occur about the line, also a certain linear tension.  With
respect to these quantities and the temperature and potentials, relations would hold analogous to those which have been
demonstrated for surfaces of discontinuity."}}  By following this prescription, exactly analogously to the derivation of
the adsorption equation (1.1), the linear adsorption equation in the form
\begin{equation} \label{eq:18}
d\tau = - \sum\limits_{i} \Lambda_i ~d\mu_i
\end{equation}
was found in earlier work \cite{Rowlinson1982/6, Widom2004}.  We shall now see that, unlike the surface adsorption
equation (1.1) for $d\sigma$, this is not invariant to the choice of location of the contact line, as it must be, and as
was verified in the remark below Eq.(2.6).  Something is missing from (3.1).

In three-phase equilibrium there is in addition to (1.2) a second Clapeyron equation [8],
\begin{equation} \label{eq:19}
\sum\limits_{i} ~(\rho_i^\beta - \rho_i^\gamma) ~d\mu_i = 0 ,
\end{equation}
where, as in Section 1, the entropy densities $s^\alpha$ etc. in the bulk phases and the temperature $T$ are included
among the $\rho_i^\alpha$ etc. and the $\mu_i$, respectively.  Because of (1.2) and (3.2) together, there are now only
$c-1$ independent $\mu_i$.

With the Gibbs adsorption equation (1.1), which holds whatever the dividing surface with respect to which the $\Gamma_i$
are defined, and the Clapeyron equations (1.2) and (3.2), it follows from (2.5) that with infinitesimal changes $d\mu_i$,

\begin{equation} 
\label{eq:20} \sum\limits_{i} ~\Lambda_i ({\bf r}) ~d\mu_i - \sum\limits_{i} ~\Lambda_i ({\bf
r}_0)~d\mu_i = -\left({\bf e}_{\alpha\beta} ~d\sigma_{\alpha\beta} + {\bf e}_{\beta\gamma} ~d\sigma_{\beta\gamma} + {\bf
e}_{\alpha\gamma} ~d\sigma_{\alpha\gamma} \right) \cdot ({\bf r} - {\bf r}_0) . 
\end{equation} 

Only if ${\bf e}_{\alpha\beta} ~d\sigma_{\alpha\beta} + {\bf e}_{\beta\gamma} ~d\sigma_{\beta\gamma} + {\bf
e}_{\alpha\gamma} ~d\sigma_{\alpha\gamma}$ were identically 0 would the left-hand side of (3.3) be 0 for all $\bf r$ and
${\bf r}_0$, so only then would $\Sigma ~\Lambda_i~d\mu_i$ be independent of the location of the contact line, as $d\tau$
is required to be.  Equivalently, from (2.6), this would require $\sigma_{\alpha\beta} ~d{\bf e}_{\alpha\beta} +
\sigma_{\beta\gamma}~d{\bf e}_{\beta\gamma} + \sigma_{\alpha\gamma} ~d{\bf e}_{\alpha\gamma}$ to be identically 0.  We
may arbitrarily hold the direction of one of the interfaces, say ${\bf e}_{\alpha\gamma}$, fixed while making the
infinitesimal changes $d\mu_i$.  Then in terms of the two orthogonal unit vectors ${\bf e}_{\alpha\gamma}$ and ${\bf
g}_{\alpha\gamma}$ and the contact angles $\alpha$ and $\gamma$, 
\begin{equation} \label{eq:21}
\begin{split}
\sigma_{\alpha\beta}~d{\bf e}_{\alpha\beta} + \sigma_{\beta\gamma}~d{\bf e}_{\beta\gamma} + \sigma_{\alpha\gamma}~d{\bf e}_{\alpha\gamma} = \left(\sigma_{\beta\gamma} \cos \gamma ~d\gamma - \sigma_{\alpha\beta} \cos \alpha ~d\alpha \right) {\bf g}_{\alpha\gamma} \\
- \left(\sigma_{\beta\gamma} \sin \gamma ~d\gamma + \sigma_{\alpha\beta} \sin \alpha ~d\alpha \right) {\bf e}_{\alpha\gamma} . \end{split}
\end{equation}
For this to vanish would require the coefficients of both ${\bf g}_{\alpha\gamma}$ and ${\bf e}_{\alpha\gamma}$ to
vanish; and so, with $\sigma_{\alpha\beta}, ~\sigma_{\beta\gamma}$, and  $d\alpha/d\gamma$ all generally finite and
non-zero, it would require that $\cos \alpha \sin \gamma + \sin \alpha \cos \gamma = 0$; or $\sin (\alpha + \gamma) = 0$;
or $\sin \beta = 0$ with $\beta = 2\pi - (\alpha + \gamma)$ the third contact angle.  Thus, the left-hand side of (3.3)
will vanish identically only when $\beta = 0$; i.e., only when the $\beta$ phase wets the interface between the $\alpha$
and $\gamma$ phases (or when any of the phases wets the interface between the other two, $\beta$ having been singled out
only by having imagined ${\bf e}_{\alpha\gamma}$ fixed).

The non-invariance of $\Sigma ~\Lambda_i ~d\mu_i$ may be illustrated with the intersecting hyperbolas in Figs.\:6 and 7. 
Let $\mu_4,\ldots, \mu_{c+1}$ be fixed, so that, with the two Clapeyron equations, there is only one independently
variable $\mu_i$, say $\mu_1$ again, as in Section 1.  Then $\Sigma ~\Lambda_i ~d\mu_i = \Lambda_{1(2,3)} ~d\mu_1$ when
$\Lambda_2 = \Lambda_3 = 0$, so the non-invariance of $\Sigma ~\Lambda_i ~d\mu_i$ would be illustrated by the
non-invariance of $\Lambda_{1(2,3)}$ as calculated from (2.12); i.e., by the dependence of $\Lambda_{1(2,3)}$ on just
which of the two or four intersections of $\Lambda_2 = 0$ with $\Lambda_3 = 0$ in Fig.\:6 or 7 is identified as ${\bf
r}_{23}$.  For the purposes of this illustration we take for the hyperbola $\Lambda_1 = 0$ the parameters
\begin{equation} \label{eq:22}
A_1/F_1 = -2/5, ~~B_1/F_1 = -5/2, ~~D_1/F_1 = -7, ~~E_1/F_1 = -1.
\end{equation}
In Fig.\:8 we show this hyperbola along with the two points $P1$ and $P2$ from Fig.\:6, and in Fig.\:9 we show the same
hyperbola (3.5) along with the four points $P1,\ldots,P4$ from Fig.\:7.

Values of the physical parameters $\rho_i^\alpha, \Gamma_i^{\alpha\beta} ({\bf r}_0)$, etc. in (2.8) that go with the
$A_i/F_i$ etc. in (2.10), (2.11), and (3.5) and are thus required for the evaluation of $\Lambda_{1(2,3)}$ from (2.12),
are in the Appendix, along with numerically accurate values of the coordinates $x,y$ of the intersections $P1$ and $P2$
in Figs.\:6 and 8 and of $P1, \ldots, P4$ in Figs.\:7 and 9.  When ${\bf r}_{23}$ in (2.12) is taken as the point $P1$ in
Figs.\:6 and 8 one finds $\Lambda_{1(2,3)} = 5.90135$ for every point ${\bf r}_1$ on both branches of the hyperbola
$\Lambda_1 = 0$, whereas when ${\bf r}_{23}$ is identified with $P2$ one finds the different value $\Lambda_{1(2,3)} =
-56.4125$ for every point ${\bf r}_1$ on $\Lambda_1 = 0$.  Similarly, when ${\bf r}_{23}$ is identified in turn with the
four points $P1, P2, P3, P4$ in Figs.\:7 and 9 one finds for $\Lambda_{1(2,3)}$ the respective values --7.17944,
--15.694, 9.73286, 136.541, each of these for every ${\bf r}_1$ on both branches of $\Lambda_1 = 0$, but differing from
each other.  The non-invariance of $\Lambda_{1(2,3)}$ in these examples, illustrating that of $\Sigma ~\Lambda_i
~d\mu_i$, is thus manifest.

What is missing from (3.1) are terms that effectively cancel the non-invariant parts of $\Sigma ~\Lambda_i ~d\mu_i$.  In
their general treatment of the thermodynamics of line tension, Boruvka and Neumann \cite{Boruvka1977} introduced
reversible work terms involving changes of the contact angles (as well as curvature terms, which we have been
systematically omitting).  We now include such terms in the infinitesimal energy change $dU$, but since they are absent
from the bulk and surface energies they emerge only in 
$dU^\ell$, the infinitesimal change in the linear excess energy. 
For length $L$ of the three-phase contact line, we thus take $dU^\ell$ to be
\begin{equation} \label{eq:23}
dU^\ell = \sum\limits_{i} \mu_i ~dN_i^\ell + \tau ~dL + L (c_\alpha ~d\alpha + c_\beta ~d\beta + c_\gamma ~d\gamma),
\end{equation}
where $N_i^\ell$ is the linear excess of component $i$ and where, following our convention, the linear excess entropy $S^\ell$ is included among the $N_i^\ell$ with $T$ the associated $\mu_i$.  It is the terms $L(c_\alpha ~d\alpha + \cdots)$ that were absent in the earlier treatments \cite{Rowlinson1982/6, Widom2004}.  Since $\alpha + \beta + \gamma = 2\pi$, only two independent combinations of the three coefficients $c_\alpha, c_\beta, c_\gamma$ occur; e.g., $c_\alpha - c_\beta$ and $c_\beta - c_\gamma$.

These coefficients depend on the location of the contact line.  As remarked earlier, different choices put the contact
line in physically different environments in the equilibrium system.  Therefore the same contact angle changes $d\alpha,
d\beta, d\gamma$ would produce different distortions in the spatial variations of the component densities, and so induce
different changes in $dU^\ell$, for different locations of the contact line. The line tension
$\tau$ is the excess grand-canonical free energy per unit length of the contact line,
\begin{equation} \label{eq:23}
L\tau = U^\ell - \sum\limits_{i} \mu_i N_i^\ell .
\end{equation}
From this and from (3.6) for $dU^\ell$, with $\Lambda_i = N_i^\ell/L$, one obtains as the linear adsorption equation
\begin{equation} \label{eq:24}
d\tau = -\sum\limits_{i} \Lambda_i ~d\mu_i + c_\alpha d\alpha + c_\beta d\beta + c_\gamma d\gamma
\end{equation}
in place of (3.1).

Another view of the physical origin of the additional terms $c_\alpha d\alpha + \cdots$ in (3.8), which makes it clear
why the terms that must be added to $-\Sigma ~\Lambda_i ~d\mu_i$ are proportional to the infinitesimal changes in the
contact angles that accompany the changes $d\mu_i$, is in the mean-field density-functional theory discussed in Section
4, below.  There it is also seen how in principle one may calculate $d\tau + \Sigma ~\Lambda_i ~d\mu_i$, i.e., $c_\alpha
d\alpha + c_\beta d\beta + c_\gamma d\gamma$, from a microscopic theory.

The invariance of $\tau$ to the choice of location of the contact line noted in Section 2 now implies that the dependence
of $c_\alpha d\alpha + \cdots$ on that choice must be just such as to cancel the contact-line dependence in $\Sigma
~\Lambda_i ~d\mu_i$ that was remarked in (3.3).  Thus, if $\Delta \bf r$ is a shift in the location of the contact line
and $\Delta c_\alpha, \Delta c_\beta$, and $\Delta c_\gamma$ are the resulting changes in the coefficients $c_\alpha$,
etc., we must have
\begin{equation} \label{eq:25}
\Delta c_\alpha ~d\alpha + \Delta c_\beta ~d\beta + \Delta c_\gamma ~d\gamma = -({\bf e}_{\alpha\beta}
~d\sigma_{\alpha\beta} + {\bf e}_{\beta\gamma} ~d\sigma_{\beta\gamma} + {\bf e}_{\alpha\gamma} ~d\sigma_{\alpha\gamma})
\cdot \Delta{\bf r} ,
\end{equation}
now with $d\alpha$ etc. and $d\sigma_{\alpha\beta}$ etc., as in (3.3) and (3.4), the infinitesimal changes in the
equilibrium contact angles and interfacial tensions that accompany the infinitesimal changes $d\mu_i$.  By (2.6), the
right-hand side of (3.9) may equally well be written $(\sigma_{\alpha\beta} ~d{\bf e}_{\alpha\beta} +
\sigma_{\beta\gamma} ~d{\bf e}_{\beta\gamma} + \sigma_{\alpha\gamma} ~d{\bf e}_{\alpha\gamma}) \cdot \Delta{\bf r}$.

Let us again fix $\mu_4, \ldots ,\mu_{c+1}$, so that from (3.8),
\begin{equation} \label{eq:26}
\frac{d\tau}{d\mu_1} = -\Lambda_{1(2,3)} + c_\alpha ~\frac{d\alpha}{d\mu_1} + c_\beta ~\frac{d\beta}{d\mu_1} + c_\gamma
~\frac{d\gamma}{d\mu_1} .
\end{equation}
Thus, it is not $-\Lambda_{1(2,3)}$ that is invariant to which of the two or four intersections of the hyperbolas
$\Lambda_2 = 0$ and $\Lambda_3 = 0$ is taken for the location of the contact line, but rather $-\Lambda_{1(2,3)} +
c_\alpha ~d\alpha/d\mu_1 + \cdots$.  This, with (3.9), implies that the combination
\begin{equation} \label{eq:27}
\Lambda_{1(2,3)} + \left(\frac{d\sigma_{\alpha\beta}}{d\mu_1} ~{\bf e}_{\alpha\beta} +
\frac{{d\sigma}_{\beta\gamma}}{d\mu_1} ~{\bf e}_{\beta\gamma} + \frac{d\sigma_{\alpha\gamma}}{d\mu_1} ~{\bf
e}_{\alpha\gamma}\right) \cdot \left({\bf r}_{23} - {\bf r}_{0} \right),
\end{equation}
with ${\bf r}_0$ any fixed location in the plane of Figs.\:3 and 4, is independent of which of the intersections ${\bf
r}_{23}$ of $\Lambda_2 = 0$ with $\Lambda_3 = 0$ is taken.

From (1.1),
\begin{equation} \label{eq:28}
\frac{{d\sigma}_{\alpha\beta}}{d\mu_1} = -\Gamma_1^{\alpha\beta} ({\bf r}_0) - \Gamma_2^{\alpha\beta} ({\bf r}_0)
~\frac{d\mu_2}{d\mu_1} - \Gamma_3^{\alpha\beta} ({\bf r}_0) ~\frac{d\mu_3}{d\mu_1} ,
\end{equation}
and similarly for $d\sigma_{\beta\gamma}/d\mu_1$ and $d\sigma_{\alpha\gamma}/d\mu_1$; and where ${\bf r}_0$, which is
arbitrary because of the invariance of $\sigma_{\alpha\beta}$ to the choice of dividing surface, may be conveniently
chosen to be the same as in (3.11) and to be the origin of the $x,y$ coordinate system in the equations (2.7) for the
hyperbolas $\Lambda_i = 0$.  The quantities $d\mu_2/d\mu_1$ and $d\mu_3/d\mu_1$ in (3.12) are expressible in terms of the
bulk-phase densities $\rho_i^\alpha, \rho_i^\beta, \rho_i^\gamma$ for $i = 1,2,3$ through the Clapeyron equations (1.2)
and (3.2) \cite{Widom2004},
\begin{equation} \label{eq:29} 
\frac{d\mu_2}{d\mu_1} = -\frac{1}{\Delta} ~\left| \begin{array} {cc} \rho_1^\alpha -
\rho_1^\beta ~~~\rho_3^\alpha - \rho_3^\beta \\ \\ \rho_1^\beta - \rho_1^\gamma ~~~~~ \rho_3^\beta - \rho_3^\gamma \\
\end{array} \right| ~~~~~\left(\rm fixed ~\mu_4, \ldots, \mu_{c+1}\right) 
\end{equation}
\begin{equation} \label{eq:30}
\frac{d\mu_3}{d\mu_1} = -\frac{1}{\Delta} ~\left| \begin{array} {cc} \rho_2^\alpha - \rho_2^\beta ~~~\rho_1^\alpha -
\rho_1^\beta \\ 
\\
\rho_2^\beta - \rho_2^\gamma ~~~~~ \rho_1^\beta - \rho_1^\gamma \\
\end{array} \right| ~~~~~\left(\rm fixed ~\mu_4, \ldots, \mu_{c+1}\right)
\end{equation}
\noindent with
\begin{equation} \label{eq:31}
\Delta = \left| \begin{array} {cc} \rho_2^\alpha - \rho_2^\beta ~~~~~\rho_3^\alpha - \rho_3^\beta \\
\\
\rho_2^\beta - \rho_2^\gamma ~~~~\rho_3^\beta - \rho_3^\gamma \\
\end{array} \right|~~ {.}
\end{equation}

We may now illustrate the invariance of (3.11) to the choice of ${\bf r}_{23}$; i.e., to the choice of which of the two
or four intersections of the hyperbolas $\Lambda_2 = 0$ and $\Lambda_3 = 0$ one takes in (3.11) when
$d\sigma_{\alpha\beta}/d\mu_1$ etc. and $d\mu_2/d\mu_1$ and $d\mu_3/d\mu_1$ are as in (3.12)-(3.15).  For this purpose we
again adopt for the hyperbolas $\Lambda_2 = 0$ and $\Lambda_3 = 0$ those in Figs.\:6 and 7 as determined by (2.10) and
(2.11), the former for a case of two intersections and the latter for a case of four; and we again take the hyperbola in
Figs.\:8 and 9 as determined by (3.5) for our example of the hyperbola $\Lambda_1 = 0$.  Now from (2.12) and
(3.12)-(3.15), and again with the numerical values quoted in the Appendix, we find for the expression (3.11) the common
value $-13.7$ when ${\bf r}_{23}$ is identified as either of the two intersections $P1$ or $P2$ in Figs.\:6 and 8, again
for every point ${\bf r}_1$ on both branches of $\Lambda_1 = 0$; and the common value $-10.65$ when ${\bf r}_{23}$ is
identified as any of the four intersections $P1, \ldots, P4$ in Figs.\:7 and 9, for every point ${\bf r}_1$ on $\Lambda_1
= 0$.  Clearly, then, it is the full expression (3.11), but, as illustrated earlier, not $\Lambda_{1(2,3)}$ alone, that
is invariant to which intersection ${\bf r}_{23}$ of $\Lambda_2 = 0$ with $\Lambda_3 = 0$ is chosen as the location of
the contact line in the definition of $\Lambda_{1(2,3)}$.

In the following Section 4 we view these matters in the context of a mean-field density-functional theory, as an example
of a microscopic theory with which one may in principle calculate the coefficients $c_\alpha$ etc. in (3.8).

\maketitle
\section{Mean-field density-functional theory}

In a general density-functional theory the line tension $\tau$ would be obtained as
\begin{equation} \label{eq:32}
\tau = \min \Bigg[\int_{A} \Psi ~da - (\sigma_{\alpha\beta} R_{\alpha\beta} + \sigma_{\beta\gamma} R_{\beta\gamma} +
\sigma_{\alpha\gamma} R_{\alpha\gamma}) \Bigg], 
\end{equation}
where $A$ is the area of the Neumann triangle in Fig.\:4, $da$ is an element of area, $R_{\alpha\beta}$ etc. are the
lengths as labeled in that figure, and $\Psi$ is a functional of the spatially varying densities $\rho_i(\bf r)$.  The
designation $``\min"$ means minimization over the  $\rho_i(\bf r)$, with $\bf r$ a vector in the plane of the figure.  It
is also to be understood that the limit in which the $R_{\alpha\beta}$ etc. become infinite, and do so proportionally to
each other, is taken in (4.1).

In a general mean-field theory, and still with the convention that the entropy density $s$ is one of the $\rho_i$ and the
temperature $T$ is the associated $\mu_i$, the functional $\Psi$ would be of the form
\begin{equation} \label{eq:33}
\Psi = F ~[\rho_1({\bf r}), \ldots, \rho_{c+1} ({\bf r}); ~~\mu_1, \ldots, \mu_{c+1}] + K,
\end{equation}
where $F$ is a local functional of the $\rho_i({\bf r})$ for given $\mu_1, \ldots, \mu_{c+1}$ while $K$ is a non-local
functional of the $\rho_i({\bf r})$ but is independent of the $\mu_i$.  In this general mean-field theory $F$ depends on
the densities $\rho_i$ and the conjugate field variables $\mu_i$ explicitly via
\begin{equation} \label{eq:34}
F ~(\rho_1, \ldots, \rho_{c+1}; ~~\mu_1, \ldots, \mu_{c+1}) = e~(\rho_1, \ldots, \rho_{c+1}) - \sum\limits_{i} ~\mu_i
\rho_i + p~(\mu_1, \ldots, \mu_{c+1}),
\end{equation}
where $e$ is the mean-field energy density as an analytic function of the densities $\rho_i$, including the unstable
parts, i.e., not reconstructed by the common-tangent-plane or convex-envelope construction; while $p$ is the equilibrium
pressure as a function of the $\mu_i$.  The terms $-\Sigma ~\mu_i \rho_i + p$ in (4.3) correspond to the subtraction of
that common tangent plane from $e(\rho_1, \ldots, \rho_{c+1})$.  Thus, $F$ is positive for all values of the $\rho_i$
except when, for the given $\mu_i$, all have their bulk-phase values $\rho_i^\alpha$ or $\rho_i^\beta$ or
$\rho_i^\gamma$, when $F$, along with its first derivatives with respect to each of the $\rho_i$, are 0.

Now let the thermodynamic state change infinitesimally via the infinitesimal changes $d\mu_i$ [only $c-1$ of which are
independent because of the two Clapeyron equations (1.2) and (3.2)].  In then calculating the resulting $d\tau$ from
(4.1) we may ignore the contributions from the changes in the minimizing $\rho_i(\bf r)$ because $\tau$ is stationary
with respect to them; we need consider only the explicit dependence of the $\Psi$ on the $\mu_i$ contributed by the terms
$-\Sigma ~\mu_i \rho_i + ~p(\mu_1, \ldots, \mu_{c+1})$ in $F$.  For the moment, we ignore the fact that the contact
angles and so also the shape of the Neumann triangle change with the $d\mu_i$, so that the region $A$ of integration in
(4.1) is for now taken to be unchanged.  Taking account of
\begin{equation} \label{eq:35}
dp = \sum\limits_{i} ~\rho_i^\alpha ~d\mu_i = \sum\limits_{i} ~\rho_i^\beta ~d\mu_i = \sum\limits_{i} ~\rho_i^\gamma
~d\mu_i
\end{equation}
from the Gibbs-Duhem equation and the Clapeyron equations (1.2) and (3.2), and of the adsorption equation (1.1), we then
have
\begin{equation} \label{eq:36}
\begin{split}
d\tau = \sum\limits_{i}  \Biggl\{ \int_{A_\alpha} \left[ -\rho_i({\bf r}) + \rho_i^\alpha \right] da + \int_{A_\beta}
\left[ -\rho_i({\bf r}) + \rho_i^\beta \right] da +  \int_{A_\gamma} \left[-\rho_i({\bf r}) + \rho_i^\gamma \right] da
\Biggr. \\
\Biggl. + ~(\Gamma_i^{\alpha\beta} R_{\alpha\beta} + \Gamma_i^{\beta\gamma} R_{\beta\gamma} +
\Gamma_i^{\alpha\gamma} R_{\alpha\gamma})  \Biggr\} d\mu_i , 
\end{split}
\end{equation}
where $A_\alpha, A_\beta, A_\gamma$ are the parts of the area $A$ as so labeled in Fig.\:4.  But the linear adsorptions
$\Lambda_i$ are
\begin{equation} \label{eq:37}
\begin{split}
\Lambda_i =  \int_{A_\alpha} \left[ \rho_i({\bf r}) - \rho_i^\alpha \right] da + \int_{A_\beta} \left[ \rho_i({\bf r}) -
 \rho_i^\beta \right] da +  \int_{A_\gamma} \left[ \rho_i({\bf r}) - \rho_i^\gamma \right] da \\ -
 ~\Gamma_i^{\alpha\beta} R_{\alpha\beta} - \Gamma_i^{\beta\gamma} R_{\beta\gamma} - \Gamma_i^{\alpha\gamma}
 R_{\alpha\gamma} , \end{split}
\end{equation}
again in the implicit limit in which the $R_{\alpha\beta}$ etc. all become infinite and do so proportionally to each
other.  Then (4.5) and (4.6) imply the (incorrect) linear adsorption equation (3.1).

Here it is clear that what is so far missing from $d\tau$ in (4.5) are the contributions from the changes in the areas of
integration $A_\alpha, A_\beta, A_\gamma$ that accompany the changes $d\mu_i$ in the thermodynamic state of the
three-phase system, and which we have so far neglected.  These changes in the areas result from changes in the contact
angles, so the missing terms in (4.5) may be taken to be of the form $c_\alpha d\alpha ~+~ c_\beta d\beta ~+~ c_\gamma
d\gamma$, as in (3.8).  With a microscopic theory such as this mean-field density-functional theory, these additional
terms may in principle be calculated because $d\tau$ and $\Sigma \Lambda_i ~d\mu_i$ are separately calculable.  For
example, with $\mu_4, \ldots, \mu_{c+1}$ again fixed, as in (3.10)-(3.15), $c_\alpha d\alpha/d\mu_1 + \cdots$ may be 
evaluated from
\begin{equation} \label{eq:38}
c_\alpha ~\frac{d\alpha}{d\mu_1} + c_\beta ~\frac{d\beta}{d\mu_1} + c_\gamma ~\frac{d\gamma}{d\mu_1} =
\frac{d\tau}{d\mu_1} + \Lambda_1 + \Lambda_2 ~\frac{d\mu_2}{d\mu_1} + \Lambda_3 ~\frac{d\mu_3}{d\mu_1} 
\end{equation}
with $d\mu_2/d\mu_1$ and $d\mu_3/d\mu_1$ obtained from (3.13)-(3.15).  Such a calculation is now being undertaken
\cite{Taylor2004}.

\maketitle
\section{Summary and conclusions}
In any plane perpendicular to a three-phase contact line the locus of choices for the location of that line that makes
the linear adsorption $\Lambda_i$ of any one of the components $i$ vanish is a rectangular hyperbola given by (2.7) with
(2.8).  Two such hyperbolas $\Lambda_j = 0$ and $\Lambda_k = 0$ may intersect in two or in four points.  The relative
adsorption $\Lambda_{i(j,k)}$, which is the adsorption of component $i$ when the contact line is so placed that the
adsorptions $\Lambda_j$ and $\Lambda_k$ both vanish, i.e., at one of the two or four intersections of $\Lambda_j = 0$
with $\Lambda_k = 0$, is given by (2.12).  Unlike its analog $d\sigma = -\Sigma ~\Gamma_i ~d\mu_i$ for an interface, the
quantity $-\Sigma ~\Lambda_i ~d\mu_i$ for a three-phase line is not invariant to the locations of the contact line and
dividing surfaces, and so cannot be the differential $d\tau$ of the necessarily invariant line tension $\tau$.  This is
illustrated in Section 3, where it is seen that $\Lambda_{1(2,3)}$ takes different values when calculated with the two or
four different intersections of the hyperbolas $\Lambda_2 = 0$ and $\Lambda_3 = 0$.  When the adsorption equation is
supplemented with terms that cancel the non-invariant parts of $\Sigma ~\Lambda_i ~d\mu_i$, in (3.8), the resulting
expression containing $\Lambda_{1(2,3)}$, in (3.11), takes the same values whichever of the intersections of $\Lambda_2 =
0$ with $\Lambda_3 = 0$ is taken for the location of the contact line, thus verifying the correctness of the modified
adsorption equation.  The added terms vanish in the limit in which one of the phases wets the interface between the other
two.

The supplemental terms in the adsorption equation arise from the changes $d\alpha, d\beta, d\gamma$ in the contact angles
accompanying the changes $d\mu_i$ in the thermodynamic state.  From a general mean-field density-functional formulation
of the calculation of line tension and the associated linear adsorptions, in Section 4, one sees explicitly why those
added terms are connected with changes in the contact angles; i.e., with the dependence on thermodynamic state of the
shape of the Neumann triangle in Fig.\:4.  Since both the line tension $\tau$ and the linear adsorptions $\Lambda_i$ are
explicitly calculable with such a density-functional theory, it would in principle allow the supplemental terms, which
are missing from the earlier formulation of the adsorption equation \cite{Rowlinson1982/6, Widom2004}, to be calculated.

\maketitle
\section*{Acknowledgment}
This work was supported by the National Science Foundation and the Cornell Center for Materials Research.

\begin{appendix}
\maketitle
\section*{Appendix}
Here we provide values for the physical parameters $\alpha, ~\gamma, ~\rho_i^\alpha, ~\Gamma_i^{\alpha\beta} ({\bf r}_0),
~\Lambda_i ({\bf r}_0)$, etc. on the right-hand sides of (2.8) that are consistent with the $A_i/F_i, ~B_i/F_i,
~D_i/F_i$, and $E_i/F_i$ in (2.10), (2.11), and (3.5), which determine the illustrative hyperbolas in Figs.\:6-9.  The
physical parameters are needed for the calculation of the values of $\Lambda_{1(2,3)}$ from (2.12) and the values of the
expression (3.11), with (3.12), all as quoted in Section 3.

All parameters and the results of all calculations are given in dimensionless terms, so each $\rho_i$ is in units of the
reciprocal of a microscopic volume, each $\sigma$ in units of a microscopic energy per unit area, $\tau$ in units of a
microscopic energy per unit length, each $\Gamma$ in units of the reciprocal of a microscopic area, and each $\Lambda$ in
units of the reciprocal of a microscopic length.

The contact angles were chosen to be
\begin{equation} {\nonumber}
~~~~~~~~~~~~~~~~~~~~~~~~~~~~~~\alpha = \gamma = 0.7\pi = 126^\circ, ~~\beta = 0.6\pi =  108^\circ.~~~~~~~~~~~~~~~~~~~~~~~~~~~~~~ (\rm A.1)
\end{equation}

\indent Of the remaining parameters on the right-hand sides of (2.8), for each $i$ six are independent: 
$\rho_i^\alpha - \rho_i^\beta, ~\rho_i^\beta - \rho_i^\gamma, ~\Gamma_i^{\alpha\beta} ({\bf r}_0),
~\Gamma_i^{\beta\gamma} ({\bf r}_0), ~\Gamma_i^{\alpha\gamma} ({\bf r}_0)$, and $\Lambda_i ({\bf r}_0)$.  These are to be
consistent, via (2.8), with the choices of the four ratios $A_i/F_i, ~B_i/F_i, ~D_i/F_i$, and $E_i/F_i$ specified in
(2.10), (2.11), or (3.5), but two may still be specified at will.  For all the calculations in the text we chose these to
be
\begin{equation} {\nonumber}
~~~~~~~~~~~~~~~~~~~~~~~~~~~~~~\Lambda_i ({\bf r}_0) = \Gamma_i^{\alpha\gamma} ({\bf r}_0) = 1 ~~~~(i = 1, 2, 3). ~~~~~~~~~~~~~~~~~~~~~~~~~~~~~~~~~ (\rm A.2)
\end{equation}
The values of the remaining parameters are now determined assuming the values already assigned to the $A_i/F_i$ etc. in
(2.10), (2.11), and (3.5), and are in Table 1.

The coordinates $x,y$ of the intersections $P1$ and $P2$ in Fig.\:6 and of the intersections $P1, \ldots, P4$ in Fig.\:7
are in Table 2.  To verify convincingly that the determinant in (2.9) vanishes requires that $P1, \ldots, P4$ be known
with great precision.  With the precision to which they are given in Table 2, MAPLE evaluates the determinant as $-2
\times 10^{-18}$.
\newpage{}

\noindent Table 1:  Numerical values of the parameters in (2.8) that are consistent with (2.10), (2.11), or (3.5), and
(A.1) and (A.2). \\ 
\\
\begin{center}
\begin{tabular}{|l|ccc|} \hline
~~~~&~~~~ For (2.10) ~~~~&~~~~ For (2.11) ~~~~&~~~~ For (3.5)\\ [1.0ex] \hline
$\rho_1^\alpha - \rho_1^\beta$ & & & 1.06866 \\ [1.0ex]
$\rho_1^\beta - \rho_1^\gamma$ & & & 2.75100 \\ [1.0ex]
$\Gamma_1^{\alpha\beta} ({\bf r}_0)$ & & & -4.32624 \\ [1.0ex]
$\Gamma_1^{\beta\gamma} ({\bf r}_0)$ & & & 4.32624 \\ [1.0ex]
$\rho_2^\alpha - \rho_2^\beta$ & -2.79046 & -2.10292 & \\ [1.0ex]
$\rho_2^\beta - \rho_2^\gamma$ & 1.41539 & 2.10292 & \\ [1.0ex]
$\Gamma_2^{\alpha\beta} ({\bf r}_0)$ & 5.25671 & 0.850651 & \\ [1.0ex]
$\Gamma_2^{\beta\gamma} ({\bf r}_0)$ & 1.5485 & 0.850651 & \\ [1.0ex]
$\rho_3^\alpha - \rho_3^\beta$ & 0.152786 & -0.057506 & \\ [1.0ex]
$\rho_3^\beta - \rho_3^\gamma$ & 0.152786 & 0.363079 & \\ [1.0ex]
$\Gamma_3^{\alpha\beta} ({\bf r}_0)$ & 0.850651 & 0.247214 & \\ [1.0ex]
$\Gamma_3^{\beta\gamma} ({\bf r}_0)$ & 0.850651 & -0.247214 & \\ [1.0ex] \hline
\end{tabular}
\end{center}

\newpage{}

\noindent Table 2:  Coordinates $x,y$ of the intersections $P1, P2$ marked in Fig.\:6 and $P1, P2, P3, P4$ marked in
Fig.\:7. \\
\\
\begin{center}
\begin{tabular}{lr|lll} \hline
For Fig.\:6, & $P1$ & $x,y$ & = & -3.355, ~-1.490 \\
& $P2$ & $x,y$ & = & 1.013, ~4.937 \\ [2.0ex]
For Fig.\:7, & $P1$ & $x$ & = & -3.951 245 182 973 155 802 3 \\
& & $y$ & = & -4.075 823 658 595 715 972 8 \\ [2.0ex]
& $P2$ & $x$ & = & 1.157 707 701 169 996 880 0 \\
& & $y$ & = & 1.529 799 699 747 754 165 2 \\ [2.0ex]
& $P3$ & $x$ & = & 0.570 879 622 026 697 738 57 \\
& & $y$ & = & -1.151 478 850 368 232 023 4 \\ [2.0ex]
& $P4$ & $x$ & = & 6.222 657 859 776 461 183 7 \\
& & $y$ & = & -6.302 497 190 783 806 169 0 \\ \hline
\end{tabular}
\end{center}
\end{appendix}

\newpage{}

\newpage{}
\section*{Figure Captions}
\hspace{0.45cm} {Fig. 1}.~~~Phases $\alpha$ and $\beta$ with the diffuse interface between them, and with a Gibbs dividing surface
chosen at an arbitrary height $z_0$.

{Fig. 2}.~~~Three phases $\alpha, \beta, \gamma$ and their interfaces $\alpha\beta$, $\beta\gamma$, and $\alpha\gamma$ meeting at a line of common contact.  The phases lie in the angles between the planar interfaces.  The dihedral angles between planes, also called $\alpha, \beta, \gamma$, are the contact angles.

{Fig. 3}.~~~Two alternative choices for the location of the three-phase contact line.  The view is along that line, which is perpendicular to the plane of the figure and so appears as a point, while the interfaces appear as lines meeting at that point.

{Fig. 4}.~~~The Neumann triangle, with sides perpendicular to the interfaces, intersecting them at the distances $R_{\alpha\beta}, R_{\beta\gamma}, R_{\alpha\gamma}$ from the contact line.  The interfaces divide the triangle into the areas $A_\alpha, A_\beta, A_\gamma$.

{Fig. 5}.~~~Unit vectors ${\bf e}_{\alpha\beta}$ etc. in the directions of the interfaces and unit vectors ${\bf g}_{\alpha\beta}$ etc. rotated counterclockwise from them by $90^\circ$.

{Fig. 6}.~~~Two rectangular hyperbolas making two intersections, marked as $P1, P2$.  Their coordinates $(x_1, y_1)$ and $(x_2, y_2)$ are given in the Appendix.

{Fig. 7}.~~~Two rectangular hyperbolas making four intersections, marked as $P1, P2, P3, P4$.  Their coordinates $(x_1, y_1),\ldots ,(x_4, y_4)$ are given in the Appendix.

{Fig. 8}.~~~Rectangular hyperbola $\Lambda_1 = 0$, with points $P1$ and $P2$ from Fig. 6.

{Fig. 9}.~~~The same rectangular hyperbola $\Lambda_1 = 0$ as in Fig. 8, with points $P1, P2, P3, P4$ from Fig. 7.
\begin{figure}[htp] 
              \begin{center}
              \leavevmode 
      	      \epsfxsize=19cm 
              \epsfbox{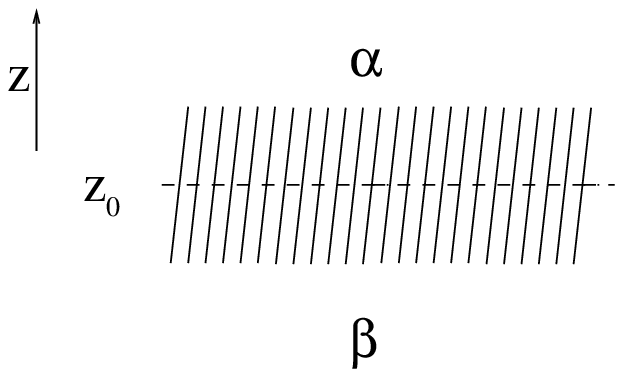}\\ \vspace{-3cm} 
              \end{center} 
              \caption{\small }
\end{figure} 
\begin{figure}[htp] 
              \begin{center}
              \leavevmode 
      	      \epsfxsize=15cm 
              \epsfbox{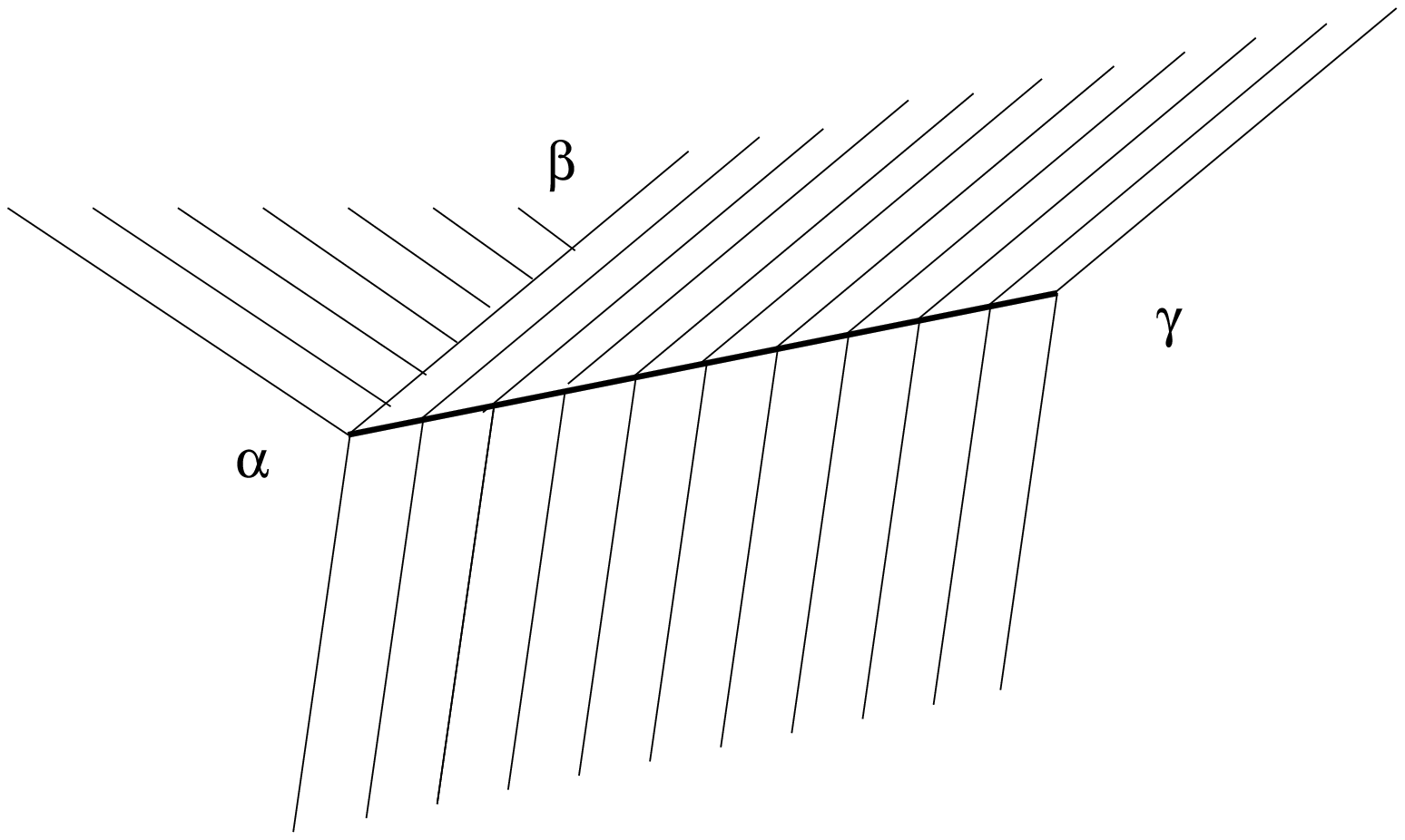}\\ \vspace{-3cm} 
              \end{center} 
              \caption{\small }
\end{figure} 
\begin{figure}[htp] 
              \begin{center}
              \leavevmode 
      	      \epsfxsize=15cm 
              \epsfbox{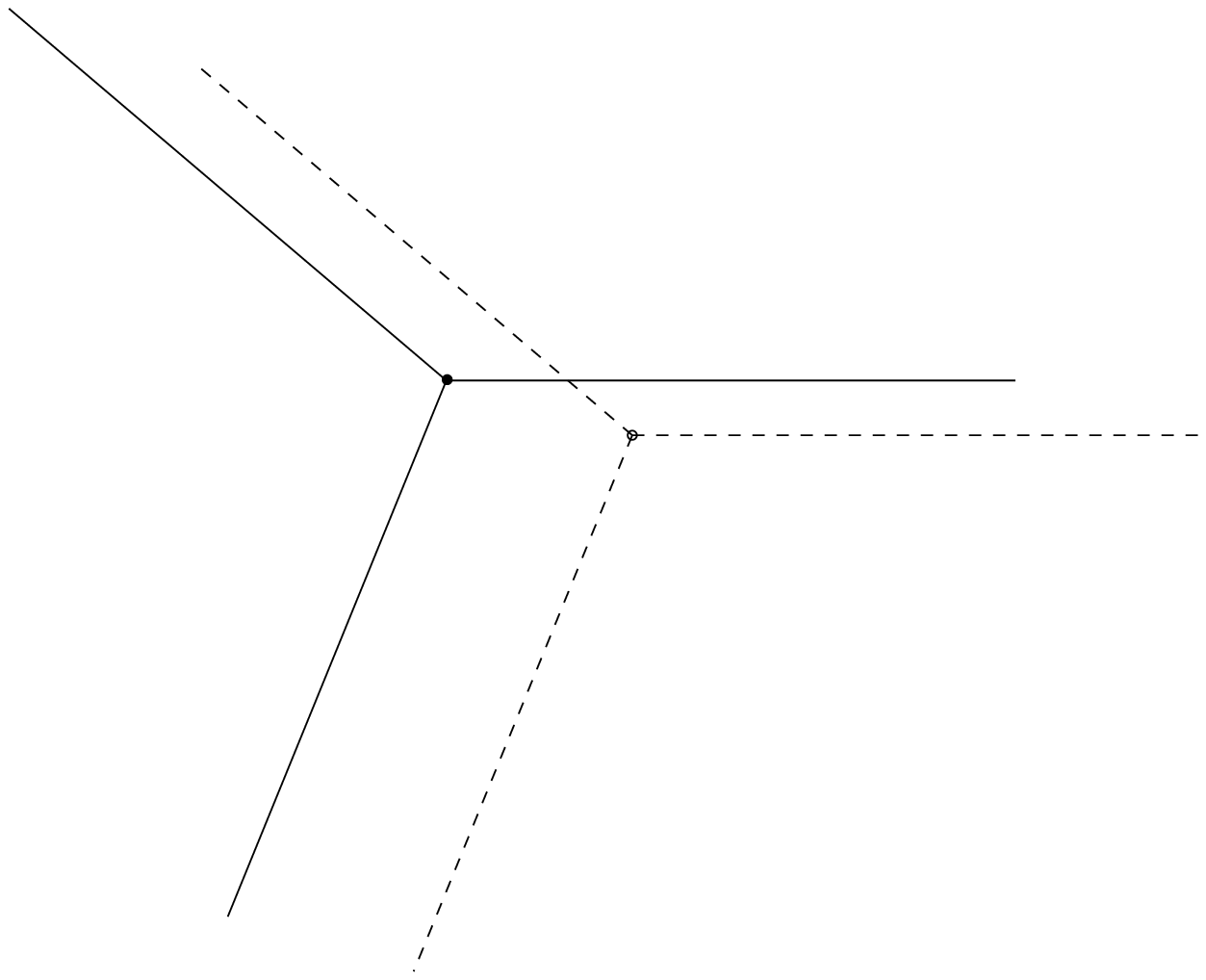}\\ \vspace{-3cm} 
              \end{center} 
              \caption{\small }
\end{figure} 
\begin{figure}[htp] 
              \begin{center}
              \leavevmode 
      	      \epsfxsize=15cm 
              \epsfbox{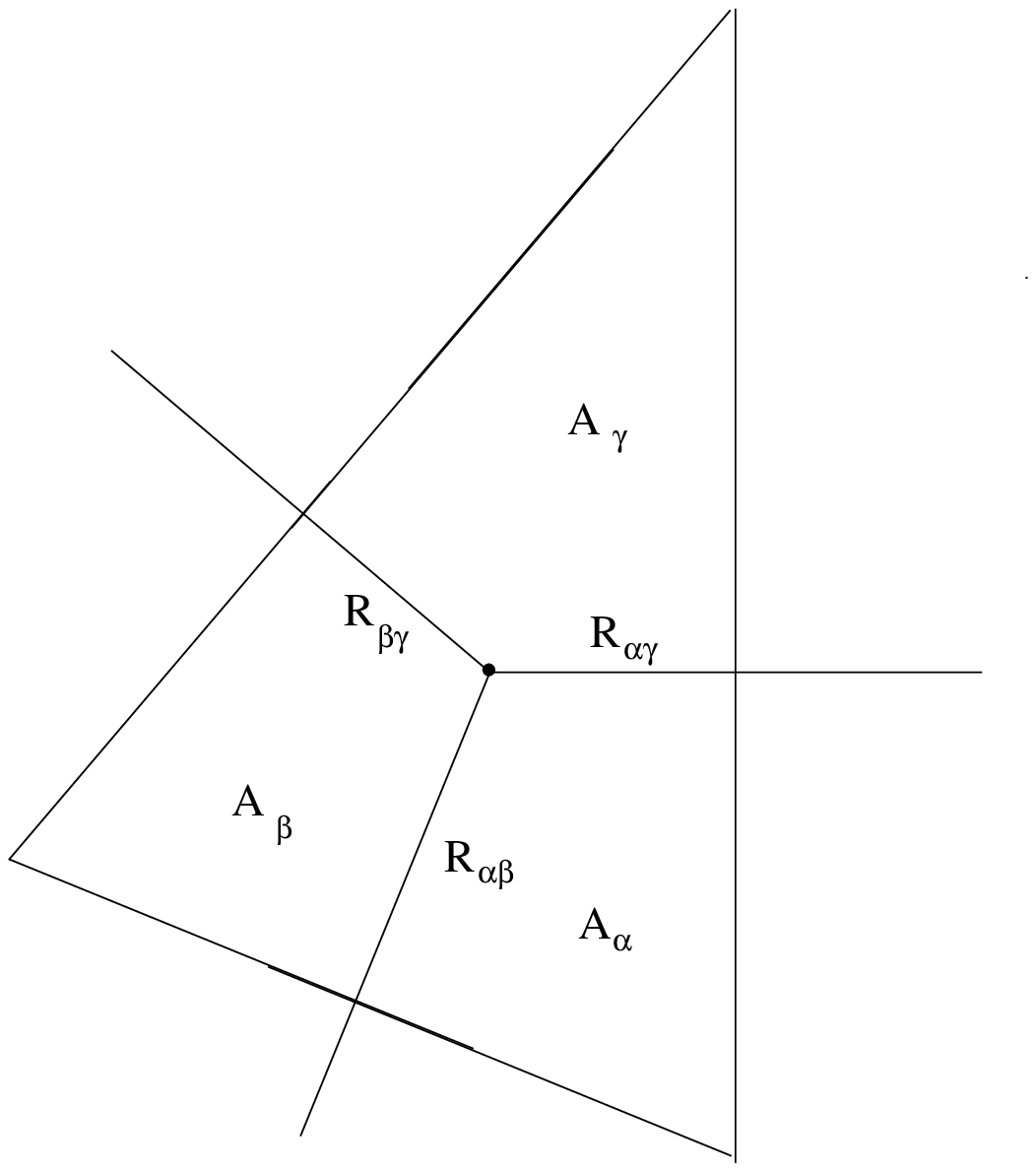}\\ \vspace{-3cm} 
              \end{center} 
              \caption{\small }
\end{figure} 
\begin{figure}[htp] 
              \begin{center}
              \leavevmode 
      	      \epsfxsize=15cm 
              \epsfbox{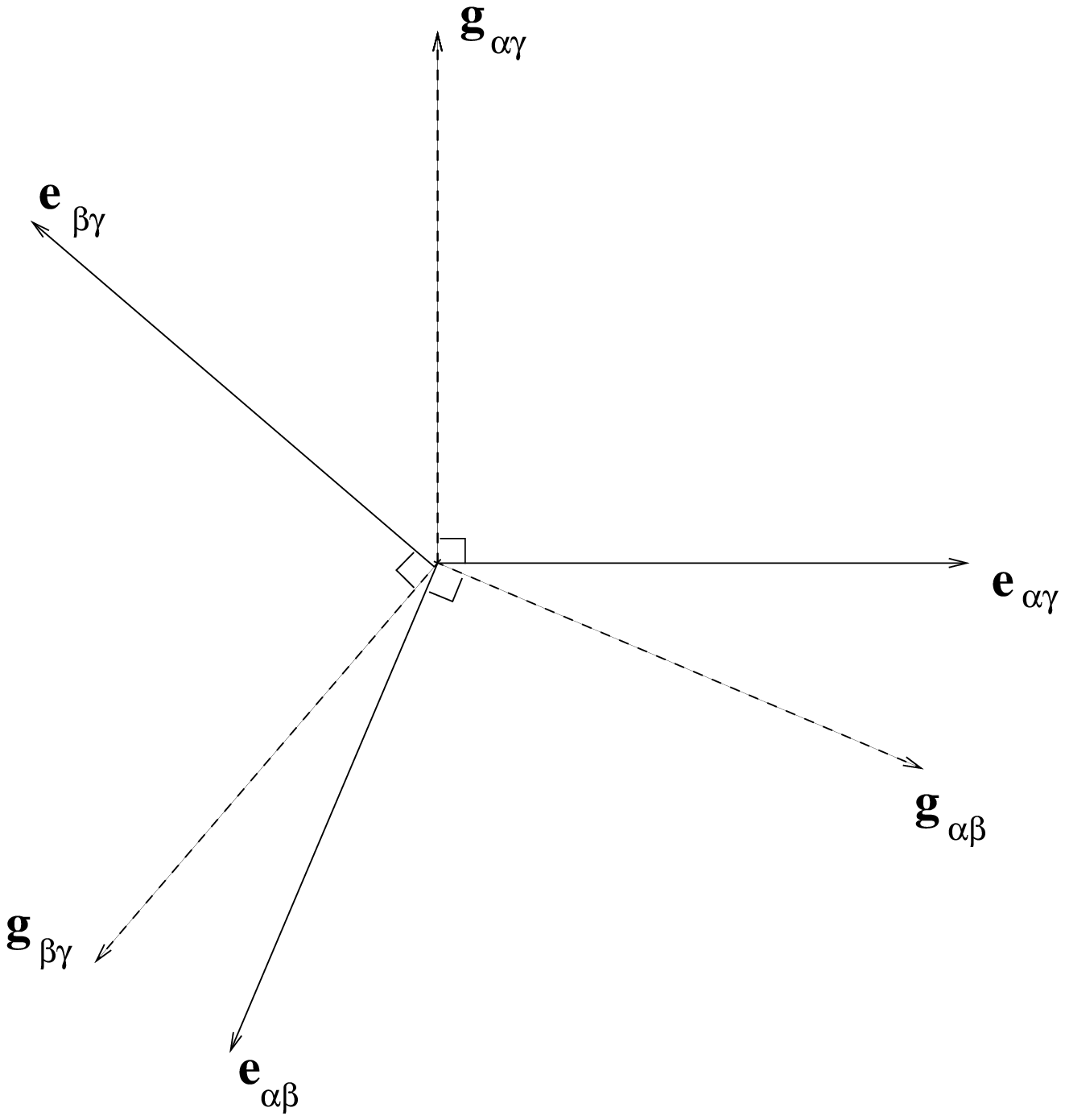}\\ \vspace{-3cm} 
              \end{center} 
              \caption{\small }
\end{figure} 
\begin{figure}[htp] 
              \begin{center}
              \leavevmode 
      	      \epsfxsize=15cm 
              \epsfbox{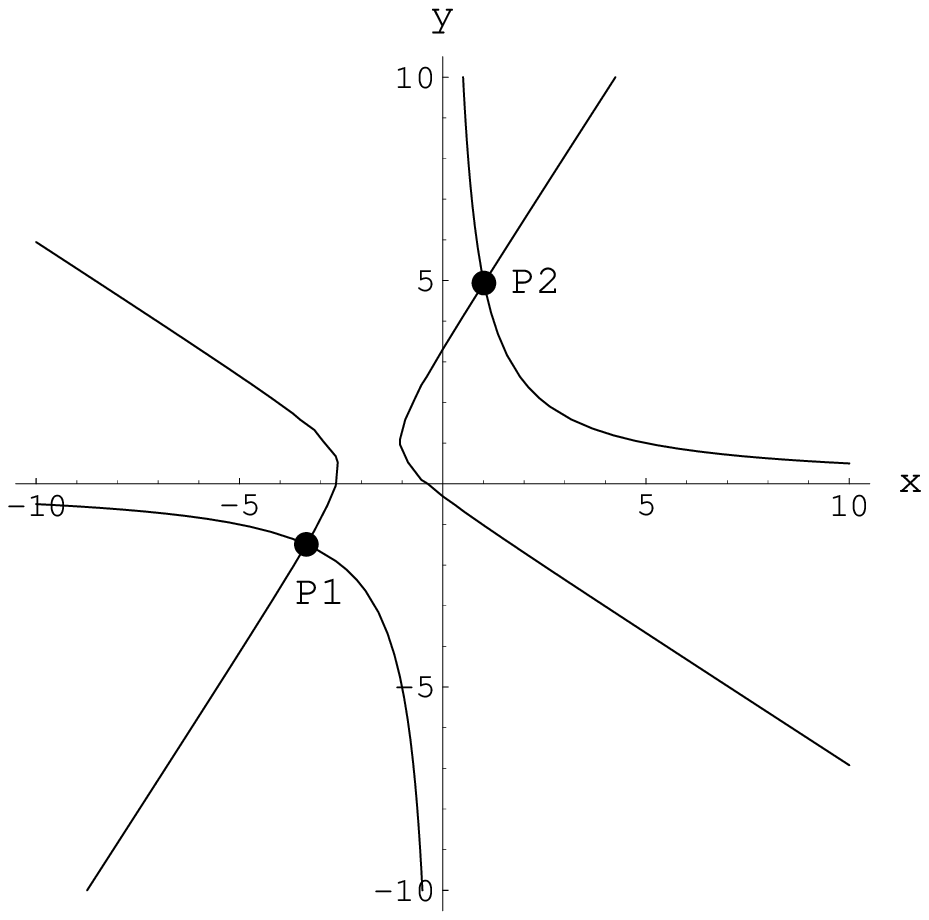}\\ \vspace{-3cm} 
              \end{center} 
              \caption{\small }
\end{figure} 
\begin{figure}[htp] 
              \begin{center}
              \leavevmode 
      	      \epsfxsize=15cm 
              \epsfbox{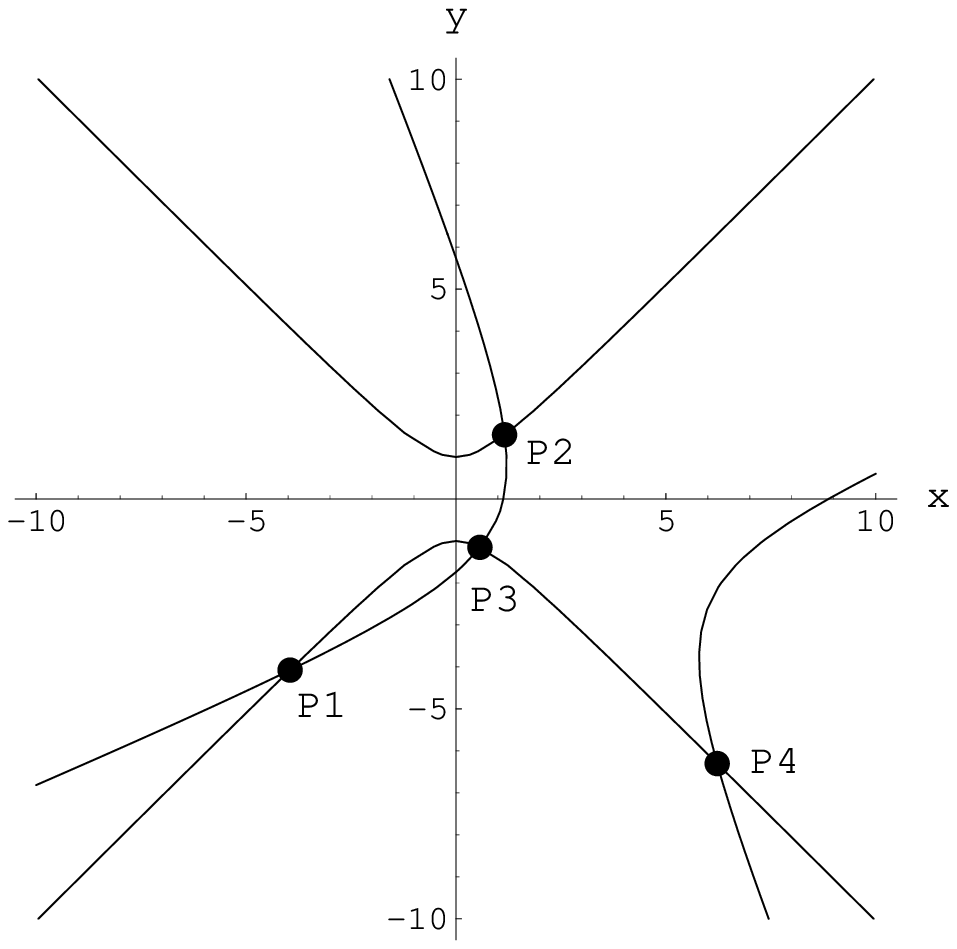}\\ \vspace{-3cm} 
              \end{center} 
              \caption{\small }
\end{figure} 
\begin{figure}[htp] 
              \begin{center}
              \leavevmode 
      	      \epsfxsize=15cm 
              \epsfbox{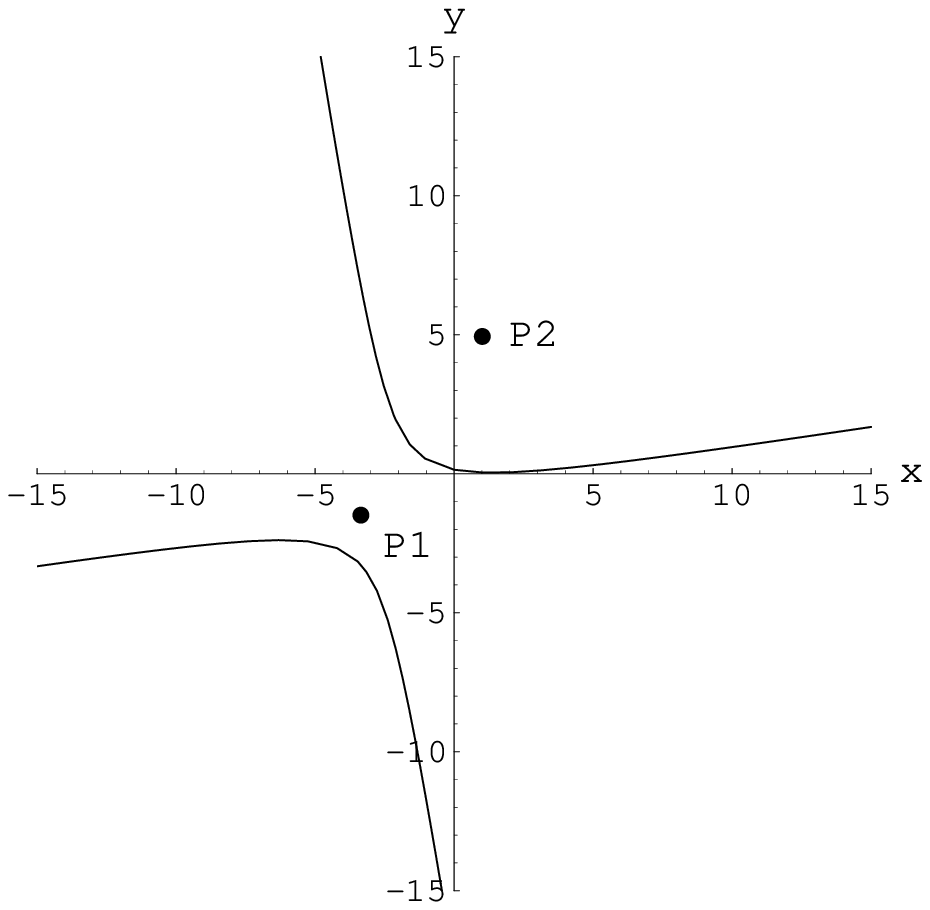}\\ \vspace{-3cm} 
              \end{center} 
              \caption{\small }
\end{figure} 
\begin{figure}[htp] 
              \begin{center}
              \leavevmode 
      	      \epsfxsize=15cm 
              \epsfbox{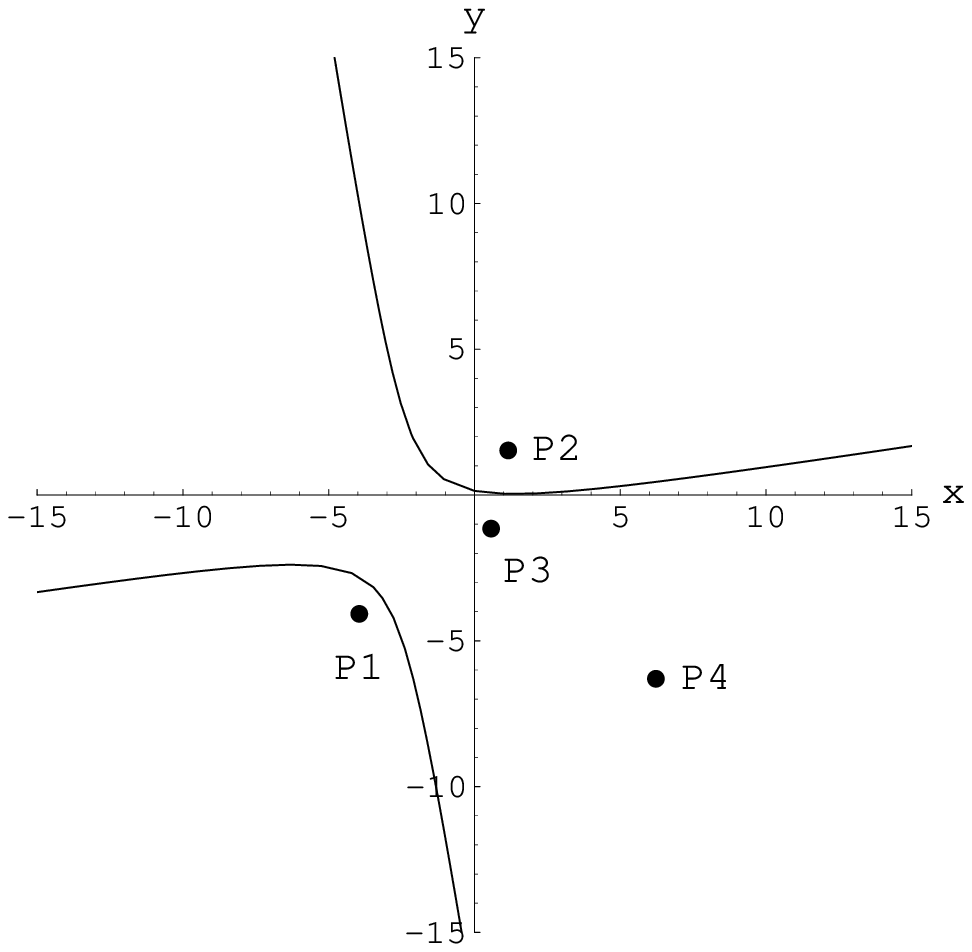}\\ \vspace{-3cm} 
              \end{center} 
              \caption{\small }
\end{figure} 


\begin{thebibliography}{11}
\bibitem{Buff1964} F.P. Buff, paper no. 1 at the 38th National Colloid Symposium, held in Austin, Texas, June 11-13, 1964.  We have not found a published version of this talk.
\bibitem{Melrose1968} J.C. Melrose, Ind. Eng. Chem. {\bf 60}, 53 (1968).
\bibitem{Good1976} R.J. Good, Pure \& Appl. Chem. {\bf 48}, 427 (1976).
\bibitem{Lovett1965} R.A. Lovett, Ph.D. thesis, U. of Rochester, 1965.
\bibitem{Rayleigh1899} Lord Rayleigh, Phil. Mag. (Series 5) {\bf 48}, 321 (1899).
\bibitem{Rowlinson1982} J.S. Rowlinson and B. Widom, {\it{Molecular Theory of Capillarity}} (Clarendon Press, Oxford, 1982), Chap. 2, Section 2.3, pp. 31-38.
\bibitem{Gibbs1928} J.W. Gibbs, {\it The Collected Works of J. Willard Gibbs}, Vol. 1 (Longmans, Green, 1928), p. 288 footnote.
\bibitem{Rowlinson1982/6} Ref. 6, Section 8.6, p. 236.
\bibitem{Widom2004} B. Widom, Coll. Surf. A (in press, 2004).
\bibitem{Boruvka1977} L. Boruvka and A.W. Neumann, J. Chem. Phys. {\bf 66}, 5464 (1977).
\bibitem{Taylor2004} C.M. Taylor, Y. Djikaev, and B. Widom, in progress.
\end{thebibliography}
\end{document}